\documentclass[12pt]{article}
\usepackage{amsfonts}
\usepackage{latexsym,amsmath}

\usepackage{amsfonts}
\usepackage{latexsym,amsmath}

\usepackage{enumerate}
\usepackage{graphicx}
\usepackage{slashbox}
\usepackage{algorithm,algorithmicx,algpseudocode}
\usepackage{enumerate}
\begin{document}
 %\pagestyle{empty}

 %An Improved Dynamic ID-based Authentication  Scheme for Telecare Medical Information Systems

\title{\bf\large The Cryptanalysis of Lee's Chaotic Maps-Based Authentication and Key Agreement Scheme using Smart card for Telecare Medicine Information Systems}%\thanks{Grants or other notes

%\author{\small Dheerendra Mishra,  Ankita Chaturvedi and  Sourav Mukhopadhyay }

\author{\small Dheerendra Mishra\thanks{E-mail:~{dheerendra@maths.iitkgp.ernet.in}}\\
\small Department of Mathematics,\\
\small Indian Institute of Technology Kharagpur,\\
\small  Kharagpur 721302, India\\}

\date{}
 \maketitle

\begin{abstract}

The Telecare medicine information system (TMIS) is developed to provide Telecare services to the remote user. A user can access remote medical servers using internet without moving from his place. Although remote user and server exchange their messages/data via  public networks. An adversary is considered to be enough powerful that he may have full control over the public network. This makes these Telecare services vulnerable to attacks. To ensure secure communication between the user and server many password based authentication schemes have been proposed. In 2013, Hao et al. presented chaotic maps-based password authentication scheme for TMIS. Recently, Lee identified that Hao et al.'s scheme fails to satisfy key agreement property, such that a malicious server can predetermine the session key. Lee also presented an efficient chaotic map-based password authentication and key agreement scheme using Smart cards for TMIS. In this article, we briefly review Lee's scheme and demonstrates the weakness of Lee's scheme. The study shows that the Lee's scheme inefficiency of password change phase causes denial of service attack and login phase results extra computation and communication overhead.

\end{abstract}
\textbf{keywords:} {Telecare medicine information system; Smart card; Password based authentication; Cryptanalysis.}

\section{Introduction}{\label{intro}}
The advancement in network technology presented and salable platform for online services. These services offer the access of server to the remote user anytime and anywhere.  Telecare medicine information systems (TMIS) are also an online healthcare service in which a patient/user can access remote medical servers. In Telecare services a remote user access the medical server via public network (Internet). An adversary may have full control over the public network, such that he can intercept, modify, delete, replay and record the messages~\cite{boyd2003protocols,xu2009improved}. This makes these services vulnerable to various kinds of attack. The vulnerability of Telecare services causes a serious threat to data security and integrity. Moreover, user privacy is always at the risk in these services.

 The adoption of authentication protocol can reduce the security risk~\cite{guerra2012authentication,aimeur2008alambic}, as a remote user and the server can authenticate each other and establish an establish a secure session. In recent years many smart card based authentication protocols have been proposed to ensure authorized medical service~\cite{awasthi2013biometric,cao2013improved,chang2013uniqueness, chen2012efficient,chen2012secure,das2013secure,das2013improved,debiao2012more,jiang2013privacy,hao2013chaotic,khan2013cryptanalysis,kumari2012cryptanalysis, lee2013efficient,lee2013secure,lin2012development,pu2012strong,tan2efficient,wei2012improved,wu2012secure,xie2013robust,yan2013secure,zhu2012efficient}. In these smart card based protocols user and server authenticate each other and establish a session key. The established session key is used for secure data transmission.

 In recent times may chaotic map-based remote user authentication scheme~\cite{guo2012chaotic,he2012cryptanalysis,lee2012extended,lee2013secure,tan2013chaotic} have been proposed to enhance the security of mutual authentication and session key agreement. In 2013, Guo and Chang~\cite{guo2012chaotic} proposed a password based authentication scheme using smart cards, which present efficient solution for smart card based authentication. In 2013, Hao et al~\cite{hao2013chaotic} proposed a smart card based password authentication scheme for  TMIS using chaotic map theory. However, Lee~\cite{lee2013efficient} shows that a malicious participant can predetermine the session key. This enables an adversary to control the communication between user and server. Moreover, Lee proposed and improved authentication scheme based on chaotic maps to ensure secure and efficient communication between user and server in TMIS. Lee claimed that his scheme is more suitable for practical TMIS. Unfortunately, Lee's scheme is vulnerable to password guessing attack and fails to present efficient login and password change phase. The inefficiency of password change phase of Lee's scheme causes denial of service attack if a user changes his password wrongly. %  where a user can never login to the server once he change

%In this article, we give a brief review of Lee's scheme and demonstrate some flaws in his scheme.

The rest of the paper is organized as follows: Section \ref{review} presents a brief review of Lee's scheme.  Section \ref{crypt} demonstrates some flaws in Lee's scheme. Finally, the conclusion is drawn in Section \ref{conclusion}.

%
%\subsection{Notations}
%
%%We will define all the notati
%\begin{table}[ht]
%  \centering
% % \caption{Meaning of symbols used throughout the paper}\label{t1}
%
%\begin{tabular}{l|l}
%%\label{table-1}
%\hline
%Notation & Descryption\\
%\hline
%$U$ & User/paitient\\
%$S$ & A trustworthy medical server\\
%$A$ & Attacker\\
%$SC$ & Smart Card\\
%$ID$ & Unique identity of $U$\\
%$PW$  & Unique password of $U$\\
%$B$ & Personal biometrics of $U$\\
%$x$  & Secret value (master key) of $S$\\
%$h(\cdot)$ & A collision resistant one-way hash function\\
%$h_1(\cdot)$  & Biohashing\\
%$\oplus$ & XOR\\
%%$\otimes$ & NOR\\
%$||$ & String concatenation operation\\
%%$N$ & Message authentication code\\
%\hline
%\end{tabular}
%
%\end{table}
\section{Review of Lin's Scheme}\label{review}
In this section, we will present the review of Lee~\cite{lee2013efficient}. This is similar to the discussion in article~\cite{lee2013efficient}.
The Lee efficient chaotic maps-based authentication scheme for Telecare medicine information system comprises the four phases, which are as follows:

\begin{itemize}
  \item Parameter generation phase
  \item Registration phase
  \item Authentication phase
  \item Password changes phase
\end{itemize}

\subsection{Notations}

%We will define all the notati
\begin{table}[ht]
  %\centering
 % \caption{Meaning of symbols used throughout the paper}\label{t1}

\begin{tabular}{l|l}
%\label{table-1}
\hline
Notation & Descryption\\
\hline

$S$ & A trustworthy remote server\\
$U$ & User\\
$ID$ & Identity of $U$\\
$PW$ & Password of $U$\\
$SC$ & Smart Card of $U$\\
$mk$ & Masterkey\\
$E_{k}(\cdot)$ & Symmetric key encryption algorithms\\
$D_{k}(\cdot)$ & Symmetric key decryption algorithms\\

$h(\cdot)$ & A one-way hash function $h: \{0, 1\}^* \rightarrow \{0, 1\}^l$\\
$h(\cdot)$ & A one-way hash function $H: [-1, 1] \rightarrow \{0, 1\}^l$\\
$\oplus$ & XOR\\
%$\otimes$ & NOR\\
$||$ & String concatenation operation\\
$\Delta T $ & Valid time delay in message transmission\\
%$N$ & Message authentication code\\
\hline
\end{tabular}

\end{table}

%%%%%%%%%%%%%%%%%%%%%%%%%%%%%%%%%%%%%%%%%%%%%%%%%%%%%%%%%%%%%%%%%%%%%%%%%%%
\subsection{Parameter generation phase}
The server $S$  chooses its master key $mk$ and generates a random number $x\in (-\infty, +\infty)$. Then, $S$ selects a one-way hash function $h(\cdot)$ and $H(\cdot)$, and secure symmetric encryption and decryption algorithms $E_{k}(\cdot)$ and $D_{k}(\cdot)$, respectively.
%%%%%%%%%%%%%%%%%%%%%%%%%%%%%%%%%%%%%%%%%%%%%%%%%%%%%%%%%%%
\subsection{Registration Phase}

A user can register his identity with server and achieve the smart card with personalized security parameters as follows:\\
\begin{description}

\item { Step 1.} $U$  selects his identity $ID$, password $PW$ and a random number $b$, then computes $W = h(PW\oplus b)$ and sends $(ID, W)$ to $S$ via secure channel.

\item { Step 2.} Upon receiving the registration request, $S$ generates a random number $r$, then computes $IM_1 = mk\oplus r$, $IM_2 = h(mk||r)\oplus ID$, $K = h(ID||mk)$ and $D_1 = h(ID||mk)\oplus h(PW\oplus b)$. Then, $S$ personalizes $U$'s smart card by embedding the parameters $\{IM_1, IM_2, D_1, h(\cdot)\}$ and provides the smart card to $U$ via secure channel.

    \item { Step 3. } Upon receiving the smart card, $U$ computes $D_2 = h(PW)\oplus b$ and stores $D_2$ into the smart card, i.e., smart card stores $\{IM_1, IM_2, D_1, D_2, h(\cdot)\}$.

\end{description}

\subsection{Authentication Phase}
In this phase $U$ and $S$ authenticate each other and draw a session key as follows:

\begin{description}
\item {Step 1.}  $U$ inserts his smart card into the card reader and inputs the password $PW$. Then, $U$ generates a random number $u$ and computes $b = D_2 \oplus h(PW)$, $K = D_1\oplus h(PW||b) = h(ID||mk)$, $T_{u}(K)$ and $X_1  = h(K||IM_1||IM_2||T_u(K)||T_1)$. Then, $U$ sends the message  $M_1 = \{IM_1, IM_2, T_u(K), X_1, T_1\}$ to $S$.

\item { Step 2. } Upon receiving $U$'s message $<M_1>$ at time $T_2$, $S$ verifies $T_2 - T_1 \leq \Delta t$, where $\Delta t$ is the valid time delay in message transmission. If the time delay in message transmission is invalid, it denies the request. Otherwise, $S$ computes $r' = IM_1\oplus mk$, $ID' = IM_2\oplus h(mk||r')$ and $K' = h(ID||mk)$, then $S$ verifies $X_1  =?~ h(K'||IM_1||IM_2||T_u(K)||T_1)$. If verification does not hold, it terminates the session. Otherwise, $S$ generates two random numbers $r_{new}$ and $v$, then computes $IM_1^* = mk\oplus r_{new}$, $IM_2^* = h(mk||r_{new})\oplus ID'$, $T_v(T_u(K))$ and $T_v(K')$. Then, $S$ computes the following values:

    $$sk = H(T_u(K), T_v(K'), T_v(T_u(K)))$$
    $$Y_1 = IM_1^*\oplus h(sk||T_2)$$
    $$ Y_2 = IM_2^*\oplus h(sk||T_2)$$
    $$Y_3 = h(sk||IM_1^*||IM_2^*||T_v(K')||T_2).$$

    Finally, $S$ sends the message $M_2 = \{Y_1, Y_2, Y_3, T_v(K'), T_2\}$ to $U$.

\item{ Step 3.} Upon receiving $S$'s message $<M_2>$ at time $T_3$, $U$ verifies $T_3 - T_2' \leq \Delta t$. If verification fails, it terminates the session. Otherwise, $U$ computes the following values $$sk' = H(T_u(K), T_v(K'), T_u(T_v(K')))$$
    $$IM_1^* = Y_1 \oplus h(sk'||T_2)$$
    $$ IM_2^* = Y_2\oplus h(sk'||T_2).$$

  Then, $U$ verifies $Y_3 =? ~ h(sk'||IM_1^*||IM_2^*||T_v(K')||T_2)$. If verification does not hold, it terminates the session. Otherwise, $U$ replaces $IM_1$ with $IM_1^*$ and $IM_2$ with $IM_2^*$.
\end{description}
\subsection{Password Change Phase}

Any legal user $U$ can change the password by adopting the following steps:

\begin{description}

\item {Step 1.} $U$ insert his smart card into the card reader and inputs his old password $PW$ and new password $PW_ {new}$ into the smart card.
\item { Step 2.} The smart card computes $b = D_2\oplus h(PW)$, $D_1' = D_1\oplus h(PW||b)\oplus h(PW_{new}||b)$ and $D_2' = h(PW_{new})\oplus b$.
\item {Step 3.} The smart card replaces $D_1$ with $D_1'$ and $D_2$ with $D_2'$.

\end{description}

\section{Analysis of Lee's Scheme}\label{crypt}  % $\{IM_1, IM_2, D_1, D_2, h(\cdot)\}$. $M_1 = \{IM_1, IM_2, T_u(K), X_1, T_1\}$ and  $M_2 = \{Y_1, Y_2, Y_3, T_v(K'), T_2\}$
In this section, we point out the flaws in Lee's scheme.

\subsection{Off-line password guessing attack}
The password guessing attacks are possible due to the following assumptions, which are discussed in various articles~\cite{brier2004correlation,boyd2003protocols,eisenbarth2008power,kocher1999differential,messerges2002examining,xu2009improved}:
\begin{itemize}
  \item In general, the user selects the passwords, which he can easily remember without storing, as long and complex passwords are usually difficult to remember and need to store. In addition, If  a user selects a long and complex password and if he forget the password, he cannot use the smart card. To avoid these difficulties, user chooses easily to remember password, which belong to a finite set.

  \item An adversary can extract the information from the smart card, $i.e.$,  an adversary can achieve $\{IM_1, IM_2, D_1, D_2, h(\cdot)\}$ from the smart card.

  \item   An adversary is able to  eavesdrop and intercept all the messages between user and server, $i.e$, he can intercept $M_1 = \{IM_1, IM_2, T_u(K), X_1, T_1\}$ and  $M_2 = \{Y_1, Y_2, Y_3, T_v(K'), T_2\}$, and record them.
\end{itemize}

With the above mentioned assumptions, a passive adversary can perform the password guessing attack as follows:

 \begin{description}
 \item {Step 1.} Guess the value $PW^*$ and compute $b^* = D_2 \oplus h(PW^*)$ and $K^* = D_1\oplus h(PW^*||b^*)$.

 \item {Step 2.} Verify $X_1  =? ~ h(K^*||IM_1||IM_2||T_u(K)||T_1)$.

 \item { Step 3.} If the verification succeeds, the password guessing succeeds. Otherwise, repeat { Step 1} and { Step 2}.
  \end{description}

\renewcommand{\labelitemii}{$-$}
\subsection{Inefficient login phase:}
A smart card based authentication scheme should be able to identify incorrect input so that extra communication and computation overhead can be avoided, as a user can input incorrect input due to mistake. However, the smart card does not very the correctness of input in Lee's scheme and executes the authentication session without verifying the correctness of input. This can be justified as follows:

 If user inputs incorrect password $PW^*$. Then without verifying the input, the smart card executes the login phase as follows:
\begin{itemize}

\item $U$ inserts his smart card into the card reader and incorrect password $PW^*$.

\item The smart card executes the session without verifying the input. It generates a random number $u$ and computes $b^* = D_2 \oplus h(PW^*)$, $K^* = D_1\oplus h(PW^*||b^*) \neq K$, as $h(PW||b) \neq h(PW^*||b^*)$.

    \item The smart card also computes  $T_{u}(K^*)$ and $X_1^*  = h(K^*||IM_1||IM_2||T_u(K^*)||T_1)$, then sends the message $M_1^* = \{IM_1, IM_2, T_u(K^*), X_1^*, T_1\}$ to $S$.

\item Upon receiving the message $<M_1^*>$ at time $T_2$, $S$ verifies $T_2 - T_1 \leq \Delta t$. The verification holds, as user usage current timestamp. Then, $S$ computes $r' = IM_1\oplus mk$, $ID' = IM_2\oplus h(mk||r')$ and $K' = h(ID||mk)$, then $S$ verifies $X_1^*  =?~ h(K'||IM_1||IM_2||T_u(K^*)||T_1)$. The verification does not hold, as $K \neq K^*$. Therefore, $S$ terminates the session.
\end{itemize}

 The inefficiency of smart card to verify the correctness of the input password causes extra computation overhead. If $T_h$, $T_{ch}$ and $T_X$ denote the time complexity of hash function, Chebyshev chaotic maps operation and XOR operation, respectively, then the  computational overheads are $4T_X + 6T_h + T_ {ch} $.

The Guo and Chang~\cite{guo2012chaotic} and Hao et al.~\cite{hao2013chaotic} scheme also do not verify the correctness of input and execute the session in case of wrong input. Therefore, both the schemes~\cite{guo2012chaotic,hao2013chaotic} also suffer this problem. 

\subsection{Inefficient password change phase}
The smart card based scheme should verify the correctness during password change. However, In Lee's scheme, the smart card executes the password change request without  verifying the correctness of input. A user may enter a wrong password, as human may sometimes forget the password or commit some mistake. This may cause the denial of service attack where a user cannot establish authorized session using the same smart card. This can be seen as follows:

  \begin{itemize}
  \item If user inputs new password $PW_{new}$ and incorrect old password $PW^*$ instead of $PW$ where $PW^* \neq PW$, the smart card does not verify the correctness of input.

   \item { Step 2.} The smart card computes $b^* = D_2\oplus h(PW^*)$, $D_1^* = D_1\oplus h(PW^*||b^*)\oplus h(PW_{new}||b^*)$ and $D_2^* = h(PW_{new})\oplus b^*$.

\item {Step 3.} The smart card replaces $D_1$ with $D_1^*$ and $D_2$ with $D_2^*$.

  \end{itemize}

The smart card inability of wrong password detection  causes \textbf{denial of service attack}, which is clear from the following points:

 \begin{itemize}

\item  $U$ inserts his smart card into the card reader and inputs the new password $PW_{new}$. Then, $U$ generates a random number $u$ and computes the following values
\begin{eqnarray*}
  b^* &=&  D_2^* \oplus h(PW_{new})\\
      &=&  h(PW_{new})\oplus b^* h(PW_{new})\\
      &=&  h(PW)\oplus b\oplus h(PW^*) \neq b, as~ h(PW) \neq h(PW^*)
\end{eqnarray*}

\begin{eqnarray*}
  K' &=& D_1^*\oplus h(PW_{new}||b^*) \\
   &=& D_1\oplus h(PW^*||b^*)\oplus h(PW_{new}||b^*)\oplus h(PW_{new}||b^*) \\
   &=& D_1\oplus h(PW^*||b^*)\\
   &=& K\oplus h(PW||b)\oplus h(PW^*||b^*) \neq ~K , as ~ h(PW||b)\neq h(PW^*||b^*)
\end{eqnarray*}

The smart card computes $T_{u}(K^*)$ and $X_1^*  = h(K^*||IM_1||IM_2||T_u(K^*)||T_1)$ and sends the message  $M_1^* = \{IM_1, IM_2, T_u(K^*), X_1^*, T_1\}$ to $S$.

\item { Step 2. } Upon receiving the message $<M_1>$ at time $T_2$, $S$ verifies $T_2 - T_1 \leq \Delta t$. The verification holds, as the user use current timestamp. Then, $S$ computes $r' = IM_1\oplus mk$, $ID' = IM_2\oplus h(mk||r')$ and $K' = h(ID'||mk)$. $S$ verifies $X_1^*  =?~ h(K'||IM_1||IM_2||T_u(K^*)||T_1)$. The verification does not hold, as $K' \neq K^*$

 \end{itemize}

This shows a user cannot login to the server if he did mistake in password change phase. 

 Hao et al.~\cite{hao2013chaotic} scheme also does not verify the correctness of input in password change phase and executes the session in case of any password input. This shows that Hao et al.~\cite{hao2013chaotic} also does not maintain efficient password change phase.

\section{Conclusion}\label{conclusion}

The presented article demonstrates the inefficiency of login and password change phase of Lee's schemes in identifying incorrect input. The study shows that flaw in password change phase leads to denial of service attack. Moreover, the study shows that Lee's scheme fails to resist password guessing attack.
  
%%%%%%%%%%%%%%%%%%%%%%%%%%%%%%%%%%%%%%%%%%%%%%%%%%%%%%%%%%%%%%%%%%%%%%%%%%%%
%
%\bibliographystyle{dheerendra}
%\bibliography{biometric-lee}

\end{document}